\documentclass[]{jfm}
   \usepackage{graphicx}
   \usepackage{amsmath,amssymb}
   \usepackage{natbib}
   \usepackage{latexsym}
   \usepackage{url}
   \usepackage{bm}
   \usepackage{color}
   \usepackage{epstopdf}
   \usepackage[margin=0cm,justification=justified]{caption}

\newcommand\Xu[1]{\textcolor{black}{#1}}
\newcommand\Song[1]{\textcolor{black}{#1}}

\title{Size-dependent transient nature of localized turbulence in transitional channel flow}
\author{Duo Xu\aff{1,2}
 \and Baofang Song\aff{3} \corresp{\email{baofang\_song@tju.edu.cn}}}

\affiliation{\aff{1}The State Key Laboratory of Nonlinear Mechanics, Institute of Mechanics, Chinese Academy of Sciences, Beijing 100190, China
             \aff{2}School of Engineering Science, University of Chinese Academy of Sciences, Beijing 100049, China
             \aff{3}Center for Applied Mathematics, Tianjin University, Tianjin 300072, China}

\begin{document}
\maketitle
\begin{abstract}
\Xu{It has been reported that a fully localized turbulent band in channel flow becomes sustained when the Reynolds number is above a threshold.}
Here we show evidences that turbulent bands are of a transient nature instead. When the band length is controlled to be fixed, lifetimes of turbulent bands appear to be stochastic and exponentially distributed, a sign of a memoryless transient nature. Besides increasing with the Reynolds number, the mean lifetime also strongly increases with the band length. Given that the band length always changes over time in real channel flow, this size dependence may translate into a time dependence, which needs to be taken into account when clarifying the relationship between  channel flow transition and the directed percolation universality class. 
\end{abstract}
\keywords{transitional channel flow, turbulent band, lifetime, size dependence}

\indent 

\section{Introduction}\label{sec:introduction}
Localized turbulence is ubiquitous in subcritical transition to turbulence in wall-bounded shear flows.
In canonical shear flows such as pipe and Couette flows, localized turbulence has been shown to be stochastically transient in nature, given that it can long live before suddenly reverting to laminar flow through a memoryless decaying process \Xu{\citep{Bottin1998, Peixinho2006, Hof2006, Eckhardt2007, Avila2010, Borrero2010, Avila2011, Shi2013}}. This transient nature is an essential ingredient (together with stochastic memoryless splitting) for establishing a relationship between the transition to turbulence in canonical shear flows and directed percolation phase transition (DP) \citep{Hinrichsen2010, Barkley2016, Lemoult2016, Goldenfeld2016, Chantry2017, Klotz2022}. These studies imply that both the transient nature and DP scenario may be universal to the transition of shear flows.

In transitional channel flow, the localized turbulence takes a form of banded structure tilted with respect to the streamwise direction \citep{Tsukahara2005}. \Song{The origin of this banded structure has been explained from the point of view of dynamical systems \citep{Paranjape2020}}. In quasi-one dimensional (1D) channel flow realized in a narrow tilted computation domain, in which turbulent bands can only localize in the direction perpendicular to them, stochastic transient and splitting behaviors have also been reported, like in other canonical shear flows \citep{Gome2020, Gome2022}. This scenario seems to suggest a (1+1)D (space + time) DP process close to the onset of sustained turbulence determined by the balance between the decay and splitting processes. However, the quasi-1D restriction does not allow the turbulent band to fully localize, leaving out the
dynamics of upstream and downstream ends of the band.

In more realistic channel flows, where turbulence can be fully localized by developing a downstream end and an upstream end, the statistical characteristic of the globally sustained turbulence was proposed to be described by the (2+1)D DP model only in a limited range of transitional Reynolds number \citep{Sano2016, Shimizu2019}. Recent studies reported that a fully localized turbulent band becomes sustained already at low Reynolds numbers $Re_\mathrm{cr}\simeq 650\sim 660$ \citep{Tao2018, Kanazawa2018, Paranjape2019, Mukund2021}, far below the critical Reynolds numbers of $Re_\mathrm{cr, DP}\simeq 830$ \citep{Sano2016} and 905 \citep{Shimizu2019} for sustained turbulence in DP theory. This early sustained turbulence conflicts with DP that turbulence should be transient and the flow should eventually laminarize completely when the Reynolds number is below $Re_\mathrm{cr,DP}$. It also renders channel flow transition distinct from its counterparts in other shear flows, challenging transition via transient turbulence as a universal scenario for shear flows.

\Xu{The self-sustaining of a turbulent band is mainly ascribed to the turbulence generation at its downstream end \citep{Kanazawa2018, Paranjape2019, Shimizu2019, Xiao2020,Xiao2020b, Liu2020, Mukund2021}.} 
\Song{\citet{Kanazawa2018} proposed that a nonlinear relative periodic orbit is embedded at the downstream end of a turbulent band, generating velocity streaks and vortices periodically,}
\Xu{while \citet{Xiao2020} and \citet{ Song2020} reported an inflectional instability associated with the local mean flow as the mechanism for turbulence generation there.}
However, a turbulent band immediately starts to decay if this mechanism is interrupted when the downstream end collides either with other turbulent bands \citep{Shimizu2019, Song2020} or with channel side wall \citep{Paranjape2019, Wu2022}. Free of disruption, a turbulent band even can be maintained down to $Re\simeq 500$, if the mechanism at the downstream end is maintained with some external body forces \citep{Song2020}. 
The dynamics at this end is thus essential for the self-sustaining of turbulent bands \citep{Mukund2021}.

In this paper, we show that the dynamics at the downstream end of a turbulent band is in fact intrinsically transient by direct numerical simulation. Unlike the localized turbulence in pipe, Couette and quasi-1D channel flows, the transient behavior of the downstream end is featured with a size dependence. 

\section{Methods}
\label{sec:methods}
\begin{figure}
\centering
\includegraphics[width=0.95\linewidth,trim={0cm 1cm 0cm 1cm}]{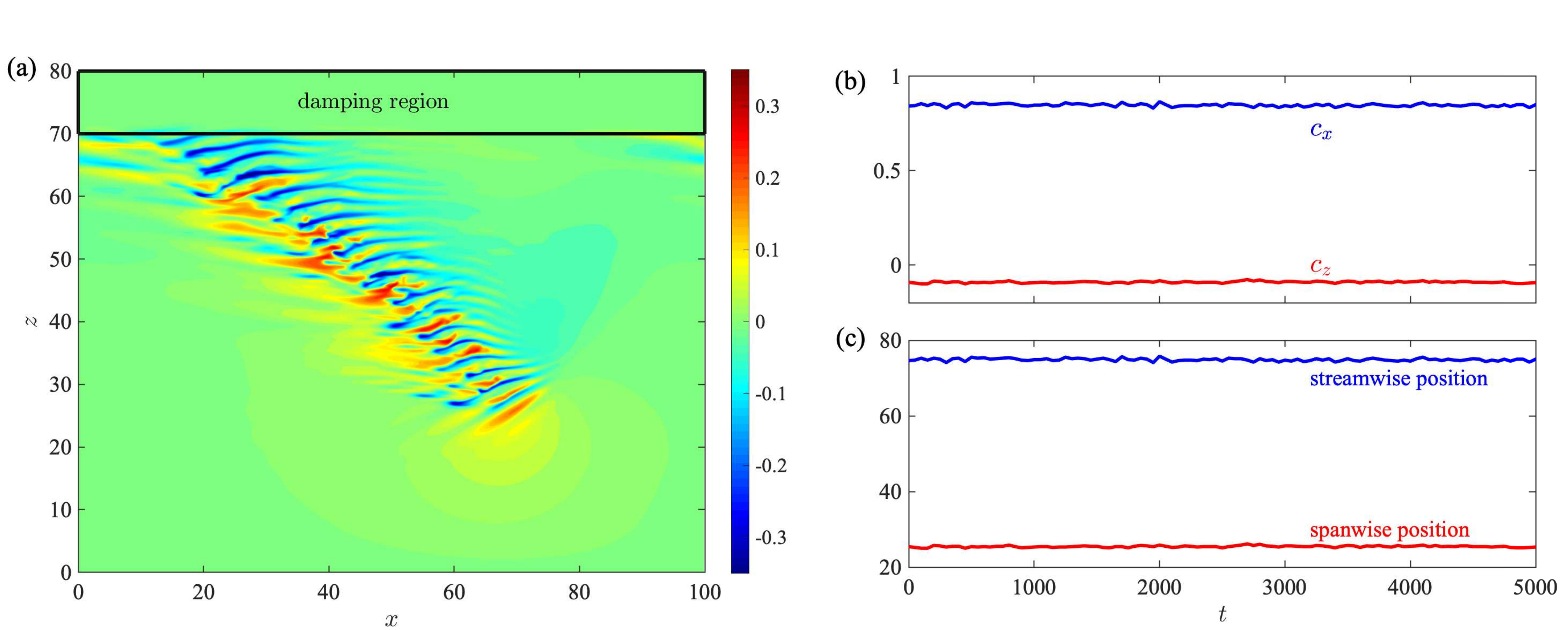}
\caption{
    \label{fig:damping_region}
     Illustration of the moving frame and damping technique used for tracking the downstream end and controlling the length of the turbulent band. (a) Contours of the fluctuations in $u_x$ are plotted. The \Xu{main flow direction} is from left to right. The downstream end is kept at approximately $(x,z)=(75,25)$ over time by dynamically adjusting the speed of the frame of reference, see panel (b). The damping region is located between $z=70$ and 80, i.e. $z_0=75$ and $R=5$ in the damping coefficient \eqref{equ:damping_coefficient}, see the rectangular region enclosed by the bold lines. With this setting, the length of the turbulent band is fixed at approximately $L=69$. (b) Time series of the adjustable speeds $c_x$ and $c_z$ of the moving frame of reference for the simulation described in panel (a). In our simulation these speeds are adjusted every 50 time units in order to keep the fluctuations in the position of the downstream end small. Reynolds number is $Re=655$. (c) The streamwise and spanwise positions of the downstream end in the moving frame described in panel (b). The fluctuations in the positions are very small (overall smaller than one length unit). Given that the position of the downstream end remains nearly unchanged during the simulations, the average speeds of the downstream end of the band can be approximated with the averaged speeds of the moving frame. In this simulation, the temporal averages are $\langle c_x\rangle=0.847$ and $\langle c_z\rangle=-0.091$, which agree with the speeds reported in the literature at close Reynolds numbers  \citep{Paranjape2019, Xiao2020b}. Note that the sign of $c_z$ is correlated with the orientation of the band.
} 
\end{figure}

It is computationally costly to track the dynamics of the entire localized bands in a large domain for a long time, given that a turbulent band may grow to hundreds of channel half heights ($h$) in length for $Re \gtrsim 660$ \citep{Kanazawa2018, Tao2018, Paranjape2019, Shimizu2019}.
Our simulations were performed in relatively smaller channels with periodic boundary conditions in the streamwise and spanwise directions, with a particular focus on the downstream end. The downstream end was tracked by using a frame of reference co-moving with the downstream end, and the band length is controlled by using a local artificial damping in a proper region of the moving frame (see figure~\ref{fig:damping_region} for an illustration). \Song{The role of this damping region is to damp out the turbulence that enters the region, \Xu{for restricting} the growth of the band. \Xu{This also helps avoiding} self-interaction of the turbulent band due to the periodic boundary condition.} This technique, \Song{inspired by \citet{Kanazawa2018},} was used to isolate turbulent fronts in short periodic pipes for successfully investigating their local dynamics \citep{Chen2022}. 

The Navier-Stokes equations in the frame of reference with a speed $\bm c=(c_x, 0, c_z)$,
\begin{equation}\label{N-S}
 \frac{\partial\boldsymbol u}{\partial t}+{(\boldsymbol u+\bm c)}\cdot\boldsymbol{\nabla}
{\boldsymbol u}=-{\boldsymbol{\nabla}p}+\frac{1}{Re}\nabla^2{\boldsymbol u} - \beta(z)(\boldsymbol u-\mathrm{\boldsymbol U}), \;
\hspace{3mm}\boldsymbol{\nabla}\cdot{\boldsymbol u}=0,
\end{equation}
are solved, where $\bm u$ is the velocity with respect to the moving frame, \Song{$\mathrm{\boldsymbol U}$ is the parabolic base flow in the moving frame}, $t$ is the time, $p$ is the pressure, and $c_x$ and $c_z$ are streamwise and spanwise speeds of the moving frame. The velocity is normalized by $3U_b/2$ and length by $h$, where $U_b$ is the bulk speed, correspondingly the Reynolds number is $Re=3U_b h/(2\nu)$ with $\nu$ the kinematic viscosity of the fluid. An artificial damping term $-\beta(z)(\boldsymbol u-\mathrm{\boldsymbol U})$ is added with
 \begin{equation}
\label{equ:damping_coefficient}
\beta(z)=A\left[1-\tanh(({|z-z_0|-R})/{B})\right],
\end{equation}
where $A$ is the nominal damping amplitude, $R$ the nominal half-width of the damping region and $B$ controls the decay of the damping strength at the boundary of the damping region ($z_0-R<z<z_0+R$). \Song{\Xu{This damping is imposed} only on the velocity deviation with respect to the parabolic base flow.} In all simulations, we chose $B=0.5$, which gives a reasonably sharp drop of the damping at the boundary of the damping region, \Song{and $A=0.25$}.
The numerical code based on a high-order finite difference and Fourier spectral schemes used by \citet{Xiao2020}, \citet{ Song2020} and \citet{Xiao2020b} is adapted for this study. The volume flux is kept constant during the simulations. The initial conditions are long-lived turbulent bands at close Reynolds numbers with the same band length, channel size and damping setting. In wall normal direction, 56 Chebyshev points are used for the finite difference discretization, and $h/\Delta x=5.2$ and $h/\Delta z=6.4$, which are comparable to the usual resolutions in the literature \citep{Kanazawa2018,Tao2018, Xiao2020, Song2020, Xiao2020b}. Time step size is $\Delta t=0.025$ for all the simulations. 
Higher wall-normal resolutions (72 and 96 points) and $\Delta t=0.015$ were also tested to make sure that the observation is not qualitatively affected.

With this technique, the position of the downstream end can be nearly fixed in the moving frame (see {figure \ref{fig:damping_region}c} for an example). Given the stable tilt angle of turbulent bands about the streamwise direction at low Reynolds numbers (around $42^\circ$ in the considered $Re$ regime) \citep{Tao2018,Kanazawa2018,Paranjape2019,
Xiao2020,Xiao2020b}, the length of the band can be well controlled.
\Xu{This allows affordable simulations and an evaluation of the dependence of the band statistics on the controlled band length.}
\Song{ With a similar idea, \cite{Taylor2016} controlled the total turbulent kinetic energy level by dynamically adjusting the Richardson number for a stratified flow.
The difference is that their control affects the flow globally, whereas our damping technique only locally affects the turbulence in a small region at the upstream end of a turbulent band.}

\section{Results}
\label{sec:results}
\begin{figure}
\centering
\includegraphics[width=0.8\linewidth, trim={1cm 0cm 2.5cm 0cm}]{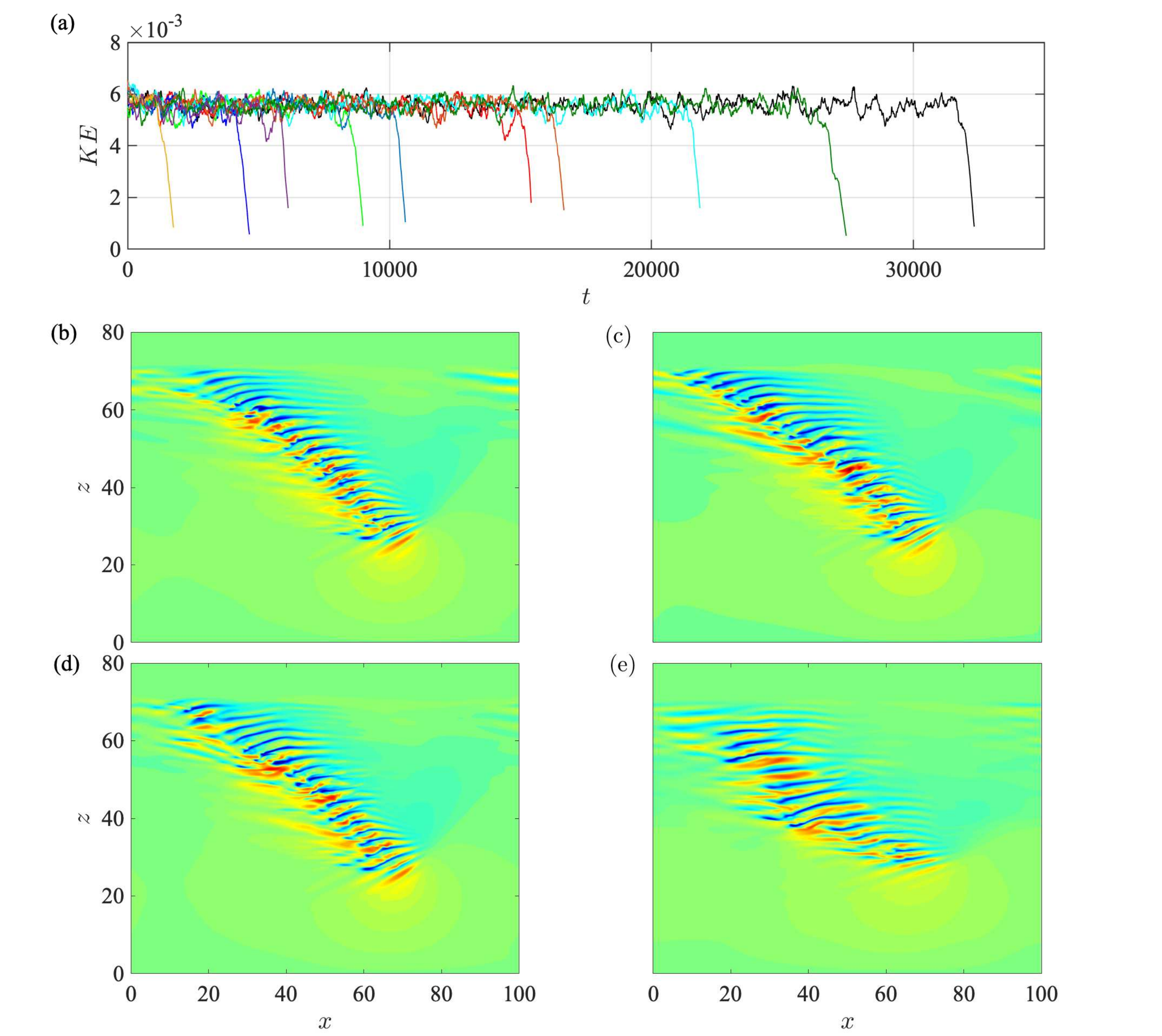}
\caption{
    \label{fig:stochastic_behavior}
     Stochastic behavior of the dynamics at the downstream end of turbulent bands at $Re=655$. The domain size is $(L_x, L_z)=(100, 80)$ and the length of the turbulent band is restricted to be approximately 69. (a) The time series of the kinetic energy of velocity fluctuations for ten runs with similar initial conditions. \Song{The initial conditions are velocity snapshots taken every 250 time units from long-lived bands at close Reynolds numbers}. (b-e) The visualization of the turbulent band in the longest run in panel (a) (the black curve) at $t=10000$, 20000, 30000, and 32050. The same quantity and color scale as those in figure \ref{fig:damping_region} are used here for the visualization. 
} 
\end{figure}

In such a setting, the downstream end can be tracked for a long time, and a stochastic transient behavior of the turbulence is observed, see  figure~\ref{fig:stochastic_behavior} for $Re=655$ (close to the reported $Re_\mathrm{cr}$).
The lengths of the bands are restricted to be around $L=69$. Figure \ref{fig:stochastic_behavior}(a) shows that the turbulence kinetic energy (\Song{$KE=\int_V(\bm u-\mathrm{\bm U})^2\mathrm{d}V/\int_V\mathrm{\bm U}^2\mathrm{d}V$ where $V$ denotes the whole flow domain}) stays at a constant level before a sharp decrease toward zero, an indication of a sudden relaminarization of the flow. No significant difference for the band is visually seen in its lifetime before the sudden decay (see figure \ref{fig:stochastic_behavior}b--d). Nevertheless, noticeable difference can be observed in the decaying process, as shown in figure \ref{fig:stochastic_behavior}(e). Firstly, the streaks at the downstream end of the band appear to be tilted at much smaller angles with respect to the streamwise direction, a possible result of a reduced inflectional instability at the downstream end \citep{Wu2022}. Secondly, the streaks away from the downstream end 
show less wiggling and fewer smaller structures, indicating weakened turbulent activities. The streak structures here cannot sustain themselves and are quickly dissipated. Once relaminarized, without a finite amplitude perturbation, the flow remains as laminar given the subcriticality of the channel flow at this low Reynolds number. \Song{See more details of the decay process in Appendix \ref{sec:decay_details}.}
 
\begin{figure}
\centering
\includegraphics[width=1\linewidth]{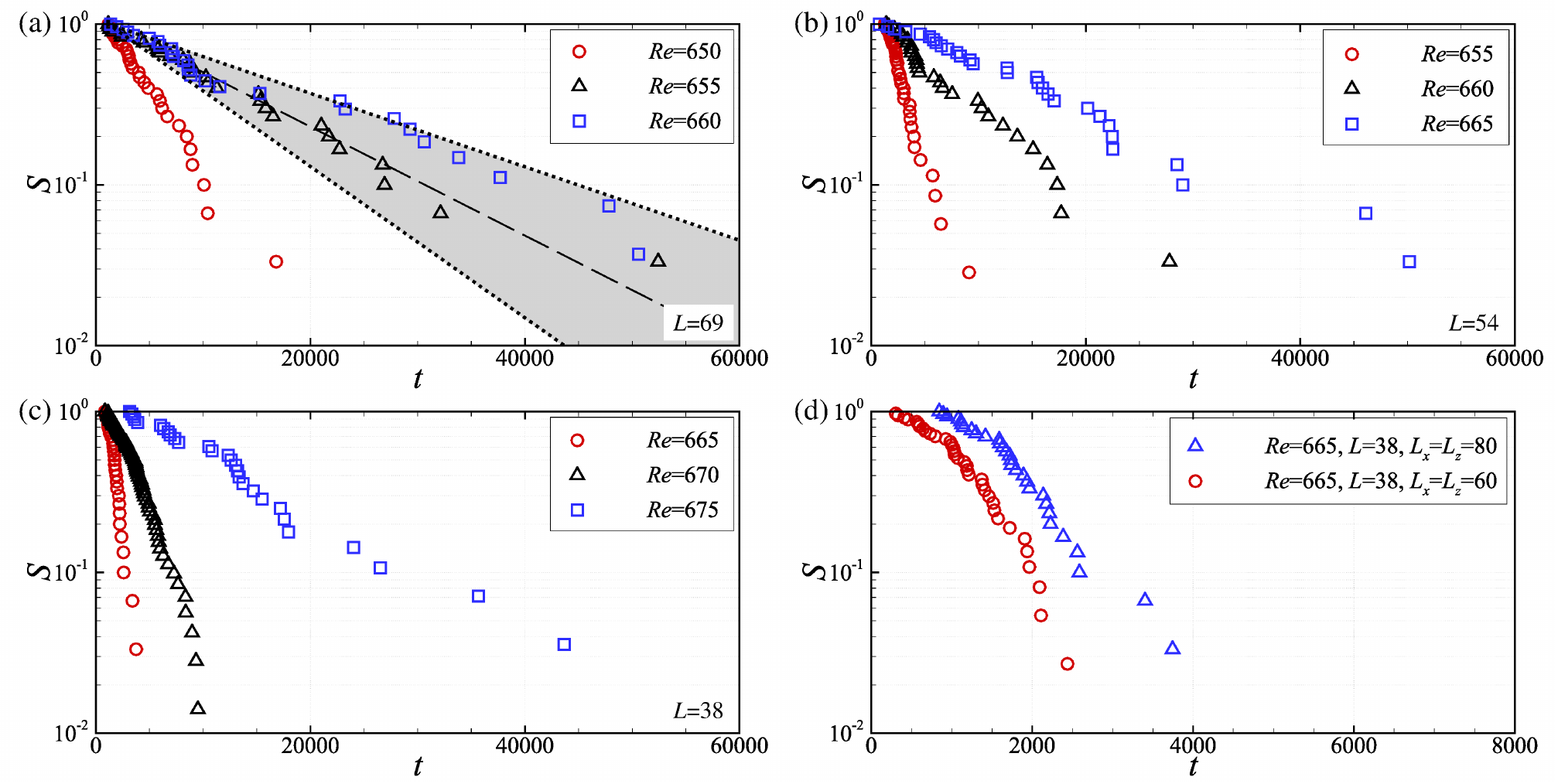}
\caption{
    \label{fig:survival_rate}
     Statistics of the survival rate $S$ of turbulent bands. $L=69$ (a), 54 (b) and 38 (c) are realized in $(L_x, L_z)=(100, 80)$, (80, 80) and (60, 60) channels, respectively. \Song{With these settings, the large-scale flow that is important for the self-sustenance of turbulent bands \citep{Duguet2013,Tao2018, Xiao2020} seems not to be significantly affected around the downstream end, see Appendix \ref{sec:large_scale_flow}}. Panels (a-c) share the $t$--axis. In (a) the dash line indicates $\mathrm{exp}[-(t-t_0)/\tau]$ where $\tau$ is the mean lifetime and $t_0$ is chosen to give minimum change of $\tau$ \citep{Gome2020}, and the shaded area between the two dotted lines denotes the $95\%$ confidence interval in $\tau$ obtained following the method in \citet{Avila2010}. In (d), the domain size effect is tested for $Re=665$ with $L=38$ in $(L_x, L_z)=(80, 80)$ and $(60, 60)$ channels. In the $(L_x, L_z)=(80, 80)$ channel, turbulent bands originally in length of $L=54$ are shortened to $L=38$ by the adjustment in the spanwise position of the damping region (see figure \ref{fig:damping_region}a). In comparison, in the $(L_x, L_z)=(60, 60)$ channel, the initial conditions are already well-developed bands with $L=38$ from $Re=675$.
The longer band-formation time due to the transient change of the band length in the former
results in the global shift in the decay time. However, no significant difference in the slopes of the two sets of data suggests that the mean lifetime is not significantly affected by the domain size if assuming an exponential distribution of the lifetime.
} 
\end{figure}

The lifetime of a band can be measured by simply setting a threshold in $KE$ ($3\times10^{-3}$ in this study), below which the turbulence is considered to have decayed. Thanks to the sharpness of the decrease, different thresholds would not cause large variation in the measured lifetime. Figure~\ref{fig:stochastic_behavior}(a) shows that the lifetimes of the bands, starting from similar initial conditions, can differ by an order of magnitude (see the yellow and black curves), given the same $Re$ and band length. This indicates the stochasticity of the lifetime.
Repeating the simulations with thirty different initial conditions at $Re=655$,
we can define the survival rate $S$ as 
\begin{equation}\label{equ:survival_rate}
S(t)=\dfrac{N_{\mathrm{survived}}(t)}{N},
\end{equation}
where $N_{\mathrm{survived}}(t)$ denotes the number of runs in which turbulent band survived up to time $t$ and $N$ denotes the sample size. Plotting $S$ against time shows that the stochastic lifetimes seem to be exponentially distributed (see the black triangles in figure~\ref{fig:survival_rate}a). The exponential distribution of the lifetime of localized turbulence has been reported in pipe, Couette, and quasi-1D channel flow \Xu{\citep{Bottin1998, Peixinho2006, Hof2006, Avila2010, Borrero2010, Shi2013, Shimizu2019b, Gome2020, Gome2022}}, however, to our knowledge, such a characteristic is for the first time observed in channel flow without the quasi-1D restriction.

Statistics of the survival rate and lifetime were then performed at a few parameter groups $(Re, L)$, as shown in figure~\ref{fig:survival_rate} where the results are grouped by the band length $L$. We considered $Re=650$, $655$ and $660$ for $L=69$, while smaller $L$'s in smaller channels for higher Reynolds numbers are considered given the rapidly increasing lifetime with $Re$ and so as the computation cost (see detailed domain configuration in the caption of figure \ref{fig:survival_rate}). 
For most $(Re, L)$ groups, we performed approximately 30 runs with similar initial conditions (limited by the high computation costs), while 75 runs were performed at $(Re, L)=(670, 38)$ considering the relatively low cost. 
The sample sizes are by no means large enough to give accurate statistical quantities, nevertheless, for all cases, we observed that the survival rate of turbulent bands $S$ exhibits an approximately exponential tail against time at sufficiently large times, \Xu{an implication for an exponential distribution.}
For each $L$, the slope of the data in semi-logarithmic scale decreases as $Re$ increases, suggesting a larger mean lifetime. 

\begin{figure}
\centering
\includegraphics[width=0.5\linewidth, trim={0cm 1cm 0cm 1.5cm}]{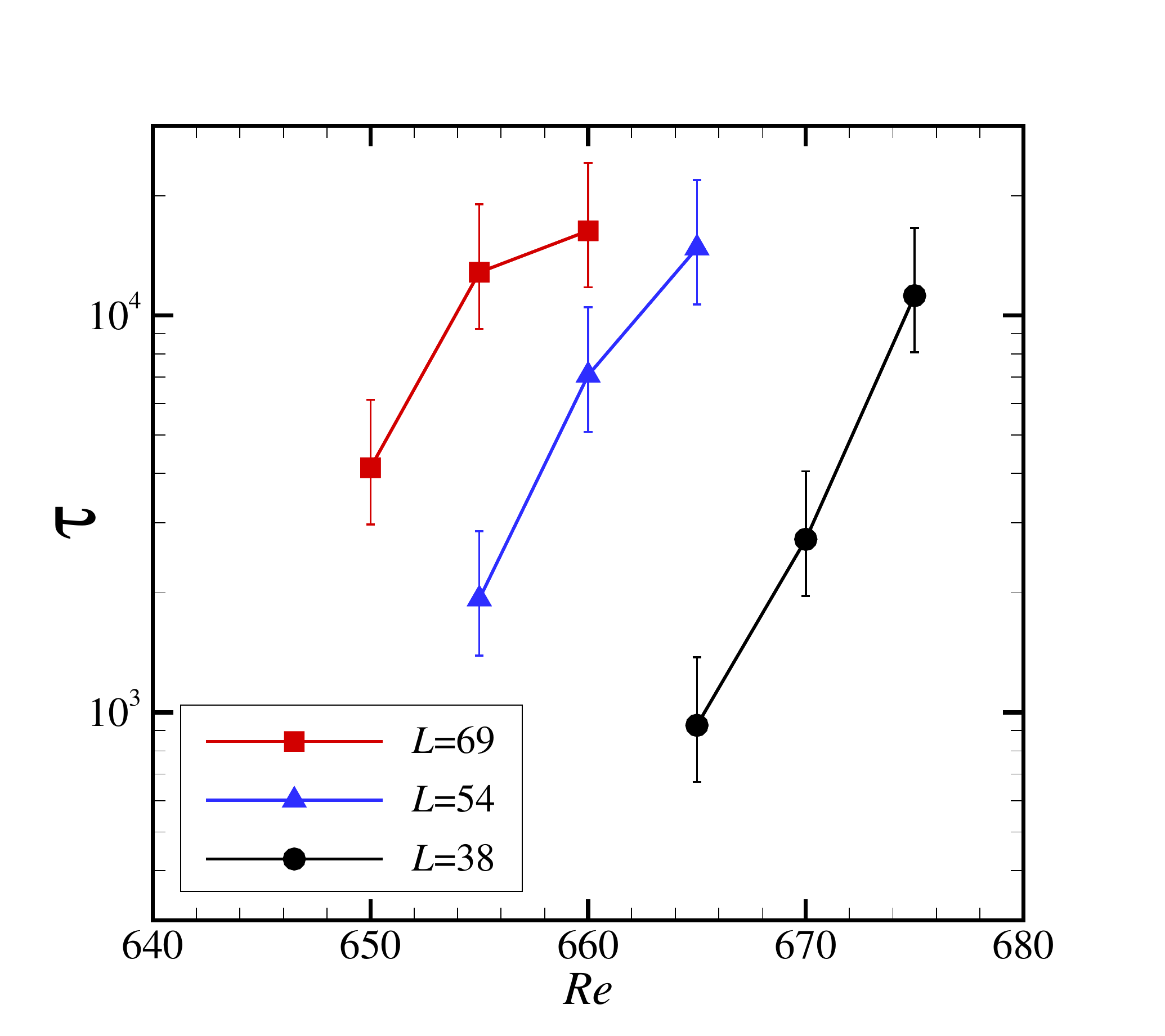}
\caption{
    \label{fig:mean_lifetime}
      \Xu{Mean lifetime $\tau$ estimated using the data shown in figure \ref{fig:survival_rate} by assuming an exponential distribution. The error bars denote the $95\%$ confidence interval of the corresponding $\tau$.}
} 
\end{figure}

Assuming an exponential distribution (suggested by figure~\ref{fig:survival_rate}), we estimated the mean lifetime (the expectation of the stochastic lifetime) of the turbulent bands using the method of \citet{Avila2010} and \citet{Gome2020}. Figure~\ref{fig:mean_lifetime} shows the mean lifetime $\tau$ against $Re$ in groups of the band length.  
Clearly, the mean lifetime increases rapidly with $Re$ in this narrow $Re$ regime, and the data for $L=38$ and $54$ seem to suggest a nearly exponential growth. However, a distinct scaling cannot be concluded due to the limited number of $Re$ and the narrow $Re$ range. 
For a fixed $Re$, a sharp increase in $\tau$ with $L$ can be seen, given that $\tau$ increases by nearly ten thousand at $Re=655$, 660 and 665 when $L$ is increased by $15$. 
Clearly, the mean lifetime strongly depends on the band length.

\begin{figure}
\centering
\includegraphics[width=0.75\linewidth]{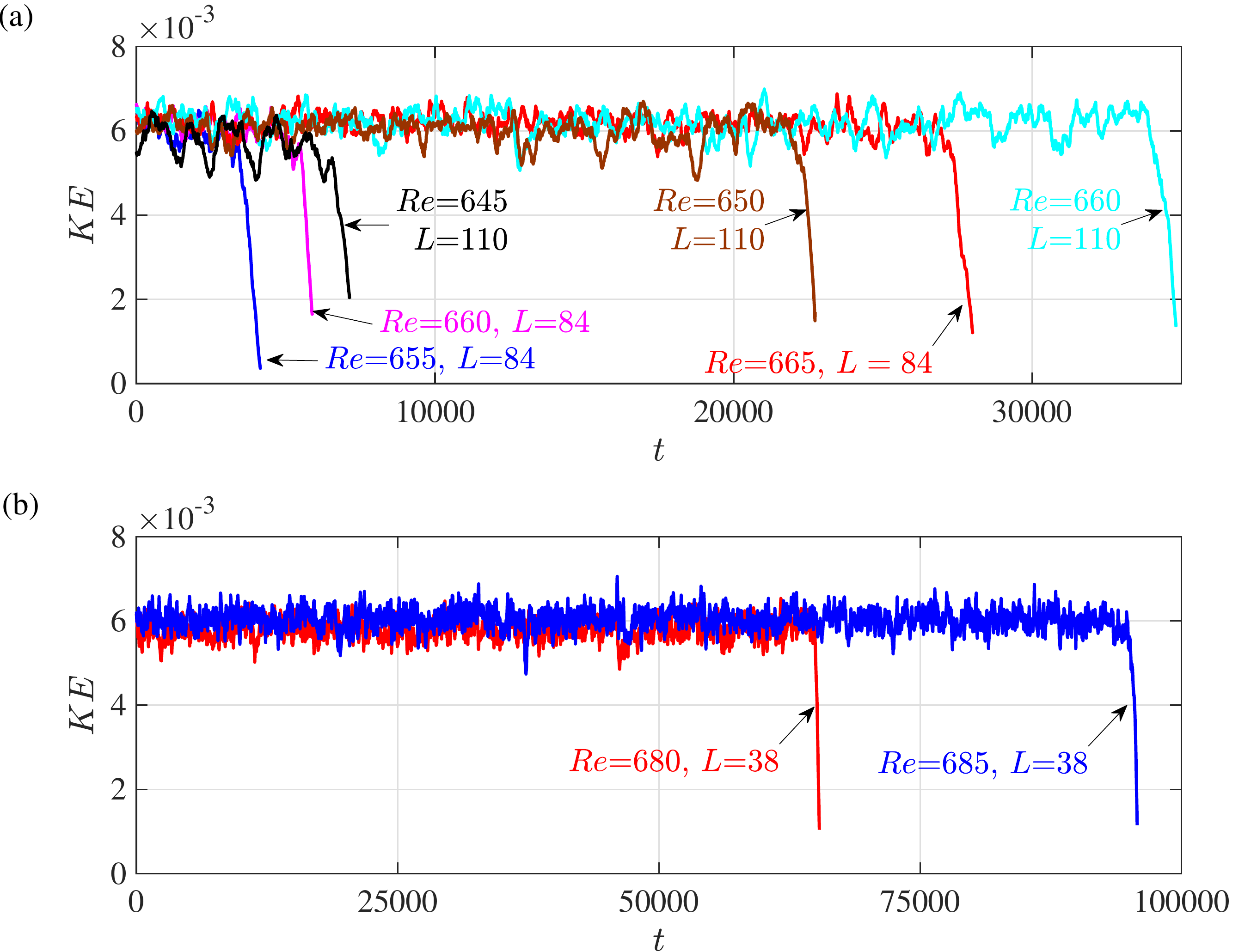}
\caption{
    \label{fig:KE_L84}
     Transient behavior at larger $L$ (a) and higher $Re$ (b). Panel (a) shows the time series of $KE$ of turbulent bands with $L=84$ realized in a channel of $(L_x, L_z)=(100, 100)$ and with $L=110$ realized in a channel of $(L_x, L_z)=(120, 110)$. Panel (b) shows $KE$ of turbulent bands with $L=38$ realized in a channel of $(L_x, L_z)=(60, 60)$ for $Re=680$ and 685.
} 
\end{figure}

Statistical studies at larger band lengths and higher Reynolds numbers are not affordable given the large length and time scales, nevertheless, we confirmed the transient behavior with a single simulation at a few parameter groups (see figure~\ref{fig:KE_L84}). \Xu{We considered $L=84$ (realized in a channel of $L_x=L_z=100$) for $Re=655$, $660$ and $665$, and also considered $L=110$ (realized in a $(L_x, L_z)=(120, 110)$ channel) for $Re=645$, $650$ and $660$.} The two lengths are about 0.3 and 0.4 times of the equilibrium length of a turbulent band at $Re=660$ measured in a large domain \citep{Kanazawa2018}, respectively. For higher $Re$, we considered $Re=680$ and 685 in a channel of $L_x=L_z=60$ with $L=38$ for the turbulent bands, and observed large lifetimes (nearly $10^5$ time units at $Re=685$), see figure~\ref{fig:KE_L84}(b). The turbulent band also shows a transient characteristic in all these simulations.
These observations give an expectation that the transient characteristic is likely to be retained beyond the band lengths and Reynolds numbers considered in the current work.

If a turbulent band is not restricted by the damping, it would grow in size before reaching the statistical equilibrium length \citep{Kanazawa2018}, and the average growth rate increases with $Re$ \citep{Paranjape2019,Mukund2021}.
Even when the band length has reached the statistical equilibrium length, it fluctuates strongly due to the erratic decay of turbulence or band splitting at the upstream end \citep{Kanazawa2018,Paranjape2019,Mukund2021}. 
This differs from turbulent puffs in pipe flow which naturally have a nearly constant size (see e.g. \citet{Wygnanski1975, Eckhardt2007, Mullin2011}), 
and therefore, the size is not relevant and the memoryless characteristic greatly eases the lifetime measurements. In channel flow, the ever change of the band length and the resulted vagueness of the transient characteristics, without any \textit{a priori} expectation of the statistical distribution of the lifetimes, pose a challenge for lifetime measurements either in experiments or simulations.

Nevertheless, our results suggest that, as the length of a turbulent band grows, the mean lifetime would increase and the relaminarization of the bands would become increasingly rare in typical observation times in experiments and most numerical simulations ($\mathcal{O}(10^3\sim 10^4)$), limited by the channel size and the computation cost. \citet{Shimizu2019} performed the longest simulations ($1.5\times 10^5$ time units), to our best knowledge, in a large domain of $(L_x, L_z)=(500, 250)$ and reported that turbulent bands only become sustained above $Re\simeq 700$. This Reynolds number is higher than the critical Reynolds number of $650\sim 660$ for sustained turbulent bands proposed in other studies \citep{Kanazawa2018, Tao2018, Paranjape2019, Kashyap2020, Mukund2021}, where the observation time was one order of magnitude shorter.
The disagreement on the critical point may be partially attributed to whether or not the transient nature of turbulent bands was observed in different studies. 

\section{Discussion and conclusion}
\label{sec:conclusion}

With a technique to control the length of a turbulent band, we identified the stochastic transient nature of the turbulence-generating dynamics at the downstream end at low Reynolds numbers, which was believed to sustain the entire turbulent band. This transient characteristic clarifies that localized turbulence in channel flow is similar to that in other shear flows and can also be described as a chaotic saddle in the perspective of dynamical systems. Our finding therefore supports that transition via transient localized turbulence is potentially a universal scenario in subcritical shear flows. Different from other flow systems, the peculiarity of {channel flow} is that, if the length of the band could be fixed, the lifetime would appear to be exponentially distributed and the decay process would be memoryless, and the mean lifetime would increase rapidly with the band length. However, when a turbulent band varies in length over time as in real channel flow, the size dependence translates into a time dependence. The  statistical distribution of the lifetime cannot be directly inferred from our results and the lifetime is even difficult to measure. Nevertheless, the transient characteristics probably change with time and an increasing mean lifetime can be expected for a growing band. 

Continuing efforts have been taken to prove that channel flow transition also belongs to the DP universality class \citep{Sano2016,Shimizu2019,Manneville2020}, thus unifying all canonical shear flows as a DP-type phase transition, which is however a difficult task given the complex dynamics exhibited by channel flow turbulence as presented here and elsewhere.
\citet{Manneville2020}
constructed a probabilistic cellular automaton model for channel flow by taking an entire turbulent band as an excited site and observed the (1+1)D DP behavior under certain model parameters.
Therefore, their study and \citet{Shimizu2019} seem to suggest a two-staged DP scenario, i.e. firstly a (1+1)D DP in the one sided regime and then a (2+1)D DP in the two sided regime.
This model requires the same time-invariant probabilistic characteristics for all sites (i.e. all bands), as the classical DP model does also, which would equivalently require the statistics of the lifetime to be the same for bands with different lengths. The size dependence of the lifetime observed in our study seems to conflict with their model assumption. 
\citet{Mukund2021} reported that the splitting of a turbulent band at the upstream end also exhibits a size-dependent stochastic behavior and proposed that the splitting is a history-dependent instead of a memoryless process, which seems to conflict with the fundamental assumption of DP also. The finding of \citet{Mukund2021} and our observation question the simplification of an entire band as a percolation site with time-invariant probabilistic characteristics.

\Song{The far-field large-scale flow around a turbulent spot was found to decay \Xu{algebraically (instead of exponentially)} away from the localized turbulence, and very large domains are needed to accommodate it given this relatively slower decay \citep{KashyapDuguetChantry2020}. Our domain sizes are not large enough in this respect. However, the  far-field velocity seems to be of low amplitudes (see figure~\ref{fig:large_scale_flow} in Appendix~\ref{sec:large_scale_flow}), and therefore is not expected to qualitatively affect the turbulence generation and self-sustenance of the band, although may have some quantitative effects on the statistics of the lifetime. This is to be studied more quantitatively in the future. Besides,} our current study covers small $Re$ and $L$ ranges so that cannot give the 
scaling of the mean lifetime with $Re$ and $L$ for bands under size control. Larger sample sizes and wider $Re$ and $L$ ranges need to be considered for future studies, and the mechanism underlying the size dependence of the lifetime remains to be theoretically clarified. The critical Reynolds number at the onset of globally sustained turbulence, above which the splitting outweighs the decay of turbulent bands,  is also an important problem to be solved in the future. It would be too costly to purely rely on brute force DNS \Song{to resolve these problems}, and more efficient algorithms, such as the extreme event algorithm \citep{Gome2022, Rolland2022} and the machine learning based method recently proposed for a simple Waleffe flow \citep{Pershin2021}, may be helpful. Simplified models may be necessary also when approaching the onset of globally sustained turbulence given the very large time and space scales involved (as in the case of pipe flow \citep{Avila2011,Barkley2016, Mukund2018}). Our finding will be informative for designing such studies.

\section{Acknowledgement}
The authors acknowledge the financial support from the National Natural Science Foundation of China under grant number 11988102, 91852105, 91752113 and 92152106, as well as from Tianjin University under grant number 2018XRX-0027. The simulations were performed on Tianhe-2 at the National Supercomputer Centre in Guangzhou and Tianhe-1(A) at the National Supercomputer Centre in Tianjin.

\section*{Declaration of interests}
The authors declare no conflicting interests.

\appendix
\section{Decay of a turbulent band}
\label{sec:decay_details}

\begin{figure}
\centering
\includegraphics[width=0.99\linewidth]{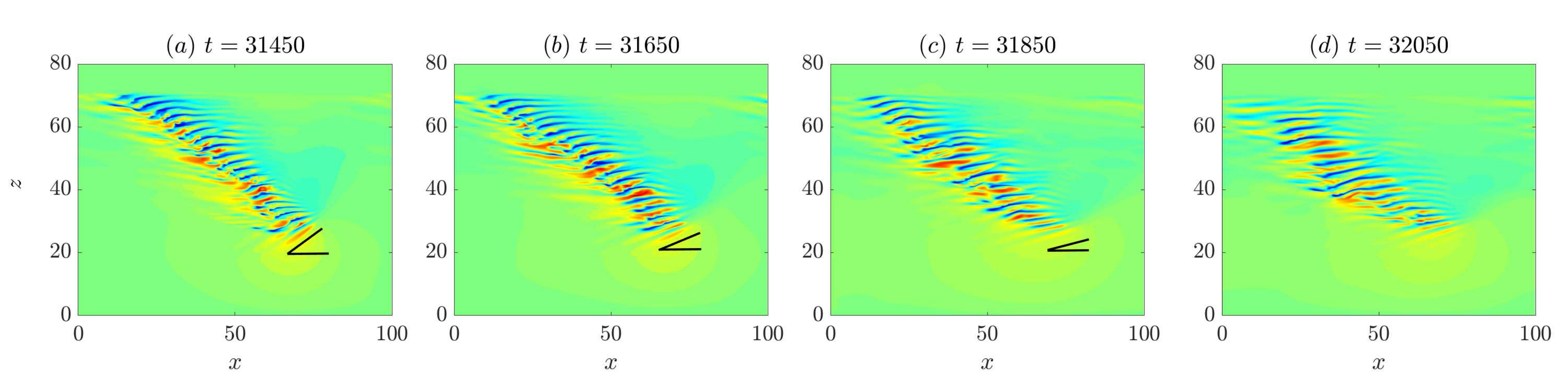}
\includegraphics[width=0.99\linewidth]{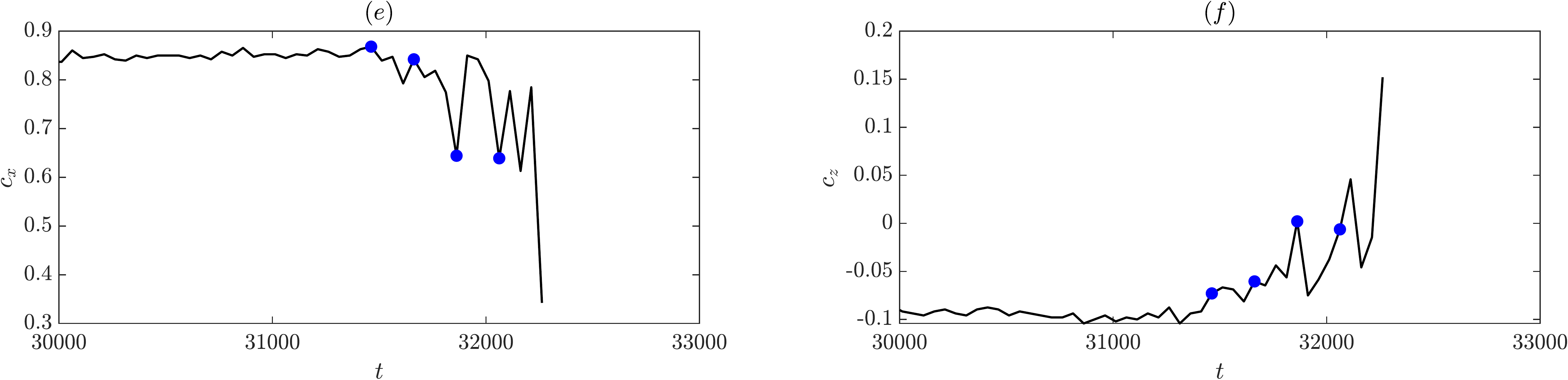}
\caption{
    \label{fig:decay_details}
     Details of the decay of the band shown in figure~\ref{fig:stochastic_behavior}(b-e). In panels (a-d), the band is visualized at time instances $t=$31450, 31650, 31850 and 32050 (the same as figure~\ref{fig:stochastic_behavior}e), which are between panels (d) and (e) of  figure~\ref{fig:stochastic_behavior}. The angle symbols are used to show the approximate tilt angles of the first high speed streak at the downstream end. In panel (e) and (f), the speeds of the frame of reference, $c_x$ and $c_z$, are plotted against time at the end of the lifetime of the band. The four blue circles mark the time instances of the corresponding panels (a-d).
} 
\end{figure}

In order to better understand the decay of the turbulence, we show more details of the decay of the turbulent band shown in figure~\ref{fig:stochastic_behavior}(b-e). The development of the velocity field at the beginning of the decay process are shown in figure~\ref{fig:decay_details}(a-d). The first sign of the decay we noticed is the decrease of the tilt angle of the streaks immediately at the downstream end, as illustrated by the angle symbols in the figure. The other sign of the decay is the significantly decreased propagation speeds $c_x$ and $c_z$ as shown in figure~\ref{fig:decay_details}(e,f). From these two signs, it can be seen that the band in figure~\ref{fig:decay_details}(a) has merely started to decay and no difference can be noticed in the flow structures compared to earlier times as shown in figure~\ref{fig:stochastic_behavior}(b-d). After 200 time units, as shown in figure~\ref{fig:decay_details}(b), the band still looks very similar to the one in figure~\ref{fig:decay_details}(a) except for the streaks immediately at the downstream end, whose tilt angle about the streamwise direction has decreased noticeably. In figure~\ref{fig:decay_details}(c), noticeable changes in both the bulk and at the upstream tail can be seen. The faint high speed streaks at the upstream edge along the band have disappeared and the downstream end has significantly decayed, indicating weakened turbulent activities. 
The streamwise and spanwise speeds of the frame of reference, which approximate the  speeds of the downstream end, are much lower than those before the decay. Particularly, the spanwise speed $c_z$ is nearly zero, indicating that the turbulence is not invading the adjacent laminar region but is passively advected by the local mean flow. At later times, $c_z$ even changes sign which indicates the turbulence is shrinking at later stages of the decay.
 \citet{Paranjape2019} and \citet{Mukund2021} reported similar phenomena for transient turbulent bands at $Re\simeq 640$. \citet{Wu2022} proposed that the decrease of the tilt angle of the streaks and the propagating speeds of the downstream end is a result of the weakened linear instability associated with the local mean flow, which was proposed to be responsible for the generation of wave-like velocity streaks and turbulence at the downstream end \citep{Xiao2020,Song2020}. Weakened linear instability gives smaller tilt angle of the flow structure and wave speed of the most unstable eigenmode, which is qualitatively consistent with the observed phenomena for a decaying band.

Another message from the figure is that the decay of the band seems to be initiated from the downstream end but not from the upstream end where the damping is acting. In other words, the decay of the turbulent band is not a direct effect of the damping but due to the transient nature of the turbulence. 
\section{The large-scale flow around a band}
\label{sec:large_scale_flow}

\begin{figure}
\centering
\includegraphics[width=0.99\linewidth]{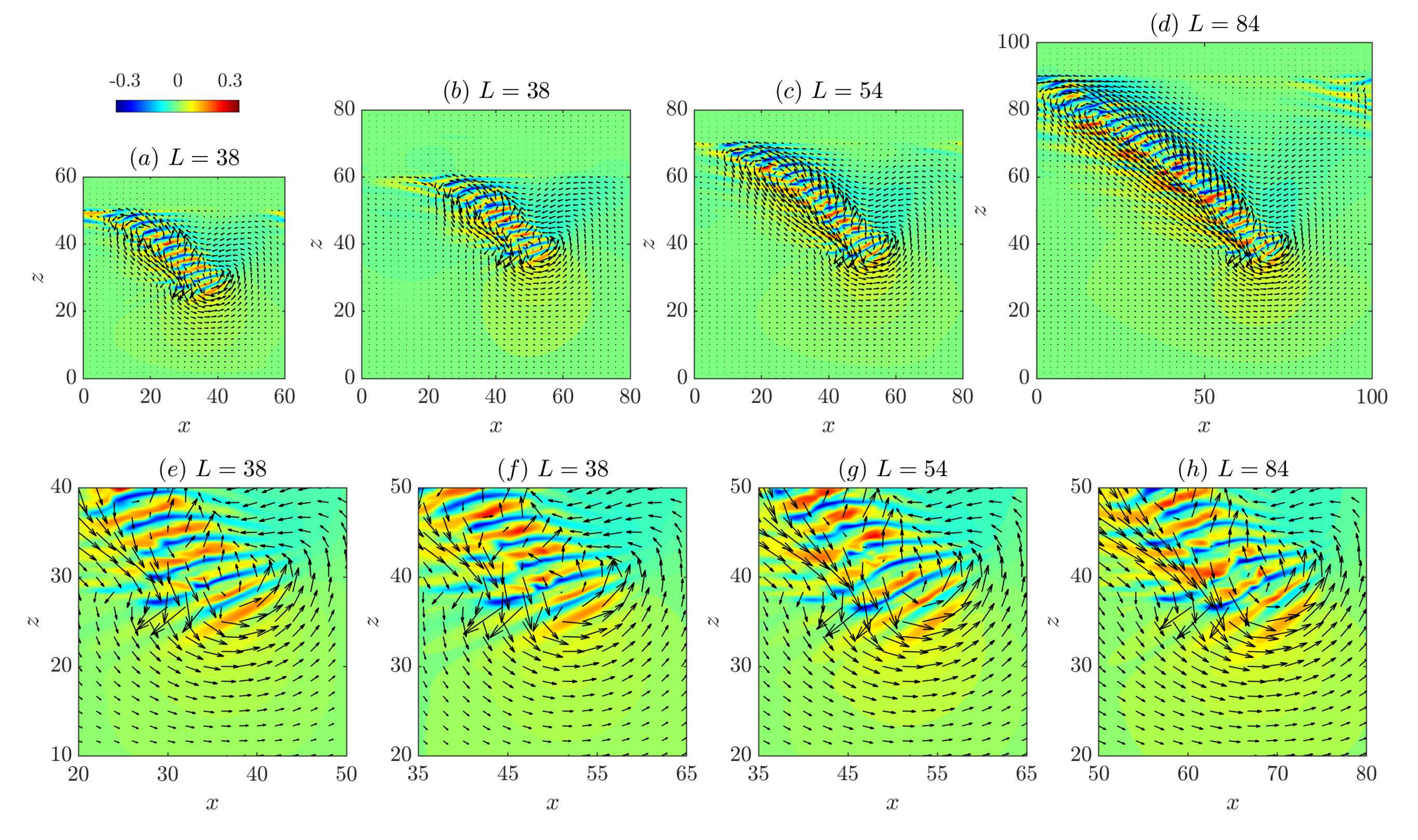}
\caption{
    \label{fig:large_scale_flow}
     The time-averaged large-scale mean flow around turbulent bands at $Re=655$ with different band lengths ($a-d$). On top of the large-scale mean flow, a snapshot of the streamwise velocity is plotted as the contours to show the band. Vectors show the in-plane velocities. The close-up view of the flow around the downstream end is shown in panels ($e-h$), corresponding to the panels ($a-d$), respectively. 
} 
\end{figure}

The large-scale flow around a band (see figure~\ref{fig:large_scale_flow}) has been considered to be important to the self-sustaining mechanism of a turbulent band in channel flow \citep{Tao2018}, plane Couette flow \citep{Duguet2013} and annular Couette flow \citep{Takeda2020}.
However, for channel flow, which part of the large-scale flow is important and how exactly it sustains the turbulent band have not been fully understood. Lately, \citet{Xiao2020} and \citet{Song2020} proposed that the part of the large-scale flow at the downstream end may be crucial, whose linear instability is responsible for the turbulence generation and the band growth. Similar mechanisms have been proposed for turbulent spots at higher Reynolds numbers \citep{Henningson1987}. The role of the large-scale flow far from the downstream end in sustaining the turbulent band remains unclear so far and is difficult to quantify, which deserves future studies.

The damping term certainly affects the local large-scale flow at the upstream end of the band, see figure~\ref{fig:large_scale_flow}(a-d). However, sufficiently far from the damping region (above roughly 10 length units away from the damping region based on the visualization), the large-scale flow visually looks the same when the band length and the domain size are increased, indicating that the damping does not noticeably affect the large-scale flow far from it, especially at the downstream end where the turbulence-generating mechanism is crucial (see figure~\ref{fig:large_scale_flow}e-h).   Similar flow pattern of the large-scale flow at the downstream end has also been reported \citep{Tao2018,Kanazawa2018, Xiao2020,Xiao2020b}. Thus, it can be inferred that the damping does not directly affect the turbulence generation at the downstream end.

\bibliographystyle{jfm}
\bibliography{references}

\end{document}